\documentclass{JINST}
\let\ifpdf\relax

\title{ An ageing study of resistive micromegas for the HL-LHC environment.}

\author{J. Gal\'an$^a$ \thanks{Corresponding author.}, D. Atti\'e$^a$, E. Ferrer-Ribas$^a$, A. Giganon$^a$, I. Giomataris$^a$, S.~Herlant$^a$, F.~Jeanneau$^a$, A.~Peyaud$^a$, Ph. Schune$^a$, T. Alexopoulos$^b$, M. Byszewski$^c$, G.~Iakovidis$^b$, P. Iengo$^d$, K. Ntekas$^b$, S. Leontsinis$^b$, R. de Oliveira$^c$, Y.~Tsipolitis$^b$, J.~Wotschack$^c$\\
\llap{$^a$}IRFU, CEA-Saclay,\\
  91191 Gif-sur-Yvette, France\\
\llap{$^b$}National Technical University of Athens,\\
Zografou Campus, GR15773, Athens, Greece\\
\llap{$^c$}CERN,\\
  Geneva, Switzerland\\
\llap{$^d$}INFN,\\
  Napoli, Italy\\
  E-mail: \email{javier.galan.lacarra@cern.ch}}

\abstract{Resistive-anode micromegas detectors are in development since several years, in an effort to solve the problem of sparks when working at high flux and high ionizing radiation like in the HL-LHC (up to ten times the luminosity of the LHC). They have been chosen as one of the technologies that will be part of the ATLAS New Small Wheel project (forward muon system). An ageing study is mandatory to assess their capabilities to handle the HL-LHC environment on a long-term period. 
A prototype has been exposed to several types of irradiation (X-rays, cold neutrons, $^{60}$Co gammas and alphas) above the equivalent charge produced at the detector in five HL-LHC running years without showing any degradation of the performances in terms of gain and energy resolution. This study has been completed with the characterization of the tracking performances in terms of efficiency and spatial resolution, verifying the compatibility of results obtained with both resistive micromegas detectors, irradiated and non-irradiated one.}

\keywords{ micromegas; ageing; HL-LHC; ATLAS }

\begin{document}

\section{Introduction}

High amplification gains are required in MicroPattern Gaseous Detectors (MPGD) in order to achieve optimum signal to noise ratios, increasing the performance of detectors in terms of spatial resolution and efficiency. The gain applied allows to observe signals from gas ionizing interactions which produce few primary electrons at the detector conversion region. The populated electron avalanches achieved with these few primary electrons entail the risk to produce a discharge at the cathode of the detector when the critical electron density of $\sim 10^7 -  10^8$ electrons per avalanche is reached (Raether limit~\cite{raether}). 

\vspace{0.2cm}

Discharges might affect the detector response in different ways; \emph{reducing its operating lifetime} due to intense currents produced in short periods of time, heating and melting the materials at the affected regions, \emph{damaging the read-out electronics} which have to support huge current loads in a brief period of time, and moreover \emph{increasing the detector dead-time} given that spark phenomena entail the \emph{discharge of the cathode} and therefore the amplification field is lost during a relatively long period of time, which is then required by the high voltage power supply to restore the charges and recover the nominal amplification field.

\vspace{0.2cm}

It was first observed in RPC-type detectors that the introduction of a high impedance resistive coating at the anode limits the detector current during a time interval of at least some microseconds, constraining the spark process to the streamer phase and reducing the total amount of charge released~\cite{fonte}. Furthermore, the limited discharge current affects the field locally, remaining in the other detector regions, and thus reducing the dead-time of the detector.

\vspace{0.2cm}

Micromegas detectors were introduced in 1995~\cite{mms} as a good candidate for high particle flux environments, and spark studies with detectors based on micromegas technology were also carried out~\cite{sparks}. Recently, additional efforts are pushing the development of resistive strip micromegas detectors in order to increase its robustness in high particle flux environments by limiting spark discharges in the same way as it was done for RPCs. In particular, the MAMMA collaboration is developing large area micromegas detectors and introduced the resistive coating technique~\cite{largeMM} for the upgrade of the HL-LHC\footnote{High Luminosity Large Hadron Collider (luminosity will be increased by at least a factor 5 reaching up to $L = 5\times10^{34}$\,cm$^{-2}$s$^{-1}$)}.

\vspace{0.2cm}

The MAMMA collaboration has investigated new detector prototypes, with different resistive coating topologies. These studies have increased the robustness and stability of micromegas detectors in the presence of an intense and highly ionizing environment, limiting the negative effect of sparks could have on them~\cite{bulk,resist}. This type of detectors will be installed in ATLAS at the New Small Wheel project. Thus, this new technology should be proved to be long term radiation resistant. The introduction of a new technology made of new materials adds the uncertainty of operation during long periods of time in intense particle flux environments. The results that we report here concern the first ageing tests with this type of detectors using different types of highly ionizing radiation.



\section{Ageing of resistive micromegas detectors.}\label{sc:ageing}

An ageing study of resistive-anodes detectors is mandatory to assess their capability to handle the rate and level of radiation at the HL-LHC. For this study, two new identical micromegas prototypes were used, lent by the MAMMA collaboration \footnote{Muon Atlas MicroMegas Activity: https://twiki.cern.ch/twiki/bin/viewauth/Atlas/MuonMicromegas} and built at the CERN workshop.

\begin{figure}[htbp]
\begin{center}
\includegraphics[width=8cm]{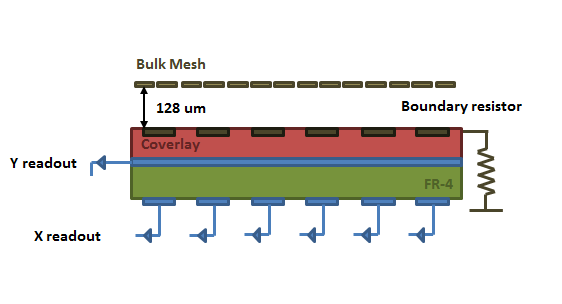}
\includegraphics[height=4cm]{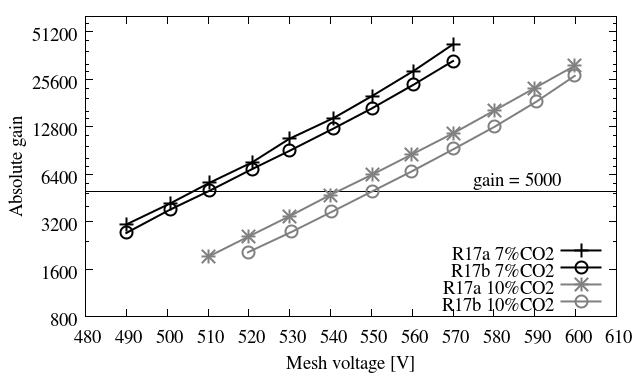}
\caption{ On the left, R17 micromegas detector schematic, including a typical 128$\mu$m amplification gap on top of resistive strips which are isolated from the X-Y copper readout strips through coverlay. On the right, gain curves of each detector at two different gas mixtures.}
\label{resist_princ}
\end{center}
\end{figure}

\vspace{0.2cm}
These detectors are based on a resistive strips technology \cite{resist} with a 2-dimensional readout. The X and Y readout strips are in copper. The top Y-strips have been covered by an insulating coverlay 60\,$\mu$m thick. The resistive strips 35\,$\mu$m thick are placed above this layer parallel to the X strips (see figure~\ref{resist_princ}).

Both, resistive and copper strips have a pitch of 250 mm and a width of 150 mm. The resistivity along the strips and boundary resistance value was measured during the fabrication process; the first detector, R17a, showed a linear resistivity of 45-50 M$\Omega$\,cm$^{-1}$, and a boundary resistance of 80-140 M$\Omega$. The resistivity obtained for the second detector, R17b, was comparable with a linear resistivity of 35-40 M$\Omega$\,cm$^{-1}$ and a boundary resistance of 60-100 M$\Omega$.

\vspace{0.2cm}
A first detector characterization in Ar+CO$_2$ mixtures \footnote{The mixtures used during ageing periods described in the text was always Ar + 10\% CO$_2$ and Ar + 7\% CO$_2$ for the test beam performances} showed the good behavior of the detectors, and the compatibility of the results obtained with both detectors, R17a and R17b. Figure \ref{resist_princ} shows the gain curves in these mixtures (amount in volume of CO$_2$ into argon).


\vspace{0.2cm}
Ageing tests took place at different CEA-Saclay facilities, one prototype (R17b) was kept unexposed as a reference and the other (R17a) was exposed to different types of radiation including X-rays, cold neutrons, high energetic gammas, and alphas, which will be described in the following sections.

\subsection{X-ray exposure}

\begin{figure}[hbp]
\begin{center}
\includegraphics[height=3.6cm]{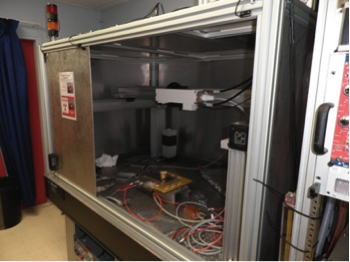}
\includegraphics[height=3.6cm]{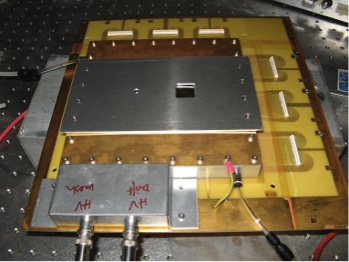}
\includegraphics[height=3.6cm]{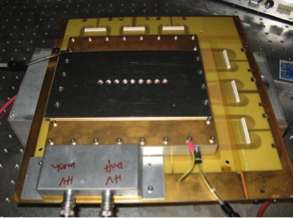}
\caption{On the left, X-ray generator cage where detector R17a was placed during irradiation. On the center, the detector prototype with the metallic mask on top where the 4 cm$^2$ squared aperture is visible together with a small hole which was closed during the irradiation tests and that could be used for calibrations at a non-exposed region. The 9-hole mask used to monitor the gain profile. }
\label{xray_setup}
\end{center}
\end{figure}

The prototype under test was placed inside a cage with a high intensity X-ray generator (see figure~\ref{xray_setup}). The X-ray generator consists of an electron gun with an accelerating power of the order of tens of kV and electron currents up to 20 mA, which points to a metallic cathode exciting and emitting X-rays, the energy of which depends on the cathode material (in our case copper with a fluorescence peak at 8 keV).

In order to study the response of the detector for a long exposure time at a specific region of the detector, a mask was prepared by using a aluminum plate with a hole of 4 cm$^2$ aperture. The plate was placed on top of the detector window, allowing a fixed position at every use (see figure~\ref{xray_setup}). This metallic irradiation mask can be placed in two different fixed positions by rotating it 180 degrees.\\

The radiation exposure tests aim to accumulate an amount of charge comparable to the values that will be integrated during the lifetime of the HL-LHC. The estimation of the total charge produced at the HL-LHC in the muon chambers of ATLAS is based on the energy deposit $E_{MIP}$ of a Minimum Ionizing Particle (MIP) in 0.5 cm micromegas conversion gap. In our gas mixture $E_{MIP} = 1.25 $ keV. Considering the detector gain $G$ at the amplification region the charge produced by each detector interaction at the HL-LHC

$$
Q_{MIP}=\frac{E_{MIP}}{W_i} q_e G
$$

where $W_i = 26.7$ eV is the mean ionization energy for the gas mixture used. Taking into account a nominal operation gain of 5000, the charge produced per MIP becomes $Q_{MIP} = 37.4$ fC.

\vspace{0.2cm}

Assuming the expected rate at the HL-LHC future muon chambers will be 10 kHz/cm$^2$ and taking 5 years of operation time (200 days/year), the total charge generated in this period will be 32.3 mC/cm$^2$. The X-ray flux produced at the X-ray generator will accumulate an amount of charge well above this value for an exposure of few days.

\vspace{0.2cm}
The detector has been exposed for more than 20 days with a gain of 5000 and a gas flow of one renewal per hour. The irradiation took place at two different operating conditions, with and without grounding the read-out strips, in order to compare the effects on the current in different conditions (further details can be found at references \cite{javier,fabien}). The current is stable, and does not show any ageing effect on the whole irradiation period, which corresponds to 21.3 days of exposure and an integrated charge of 918 mC, that is 5 years of HL-LHC with a security factor more than 1.5. 

\begin{figure}[ht!]
\begin{center}
\includegraphics[width=15cm]{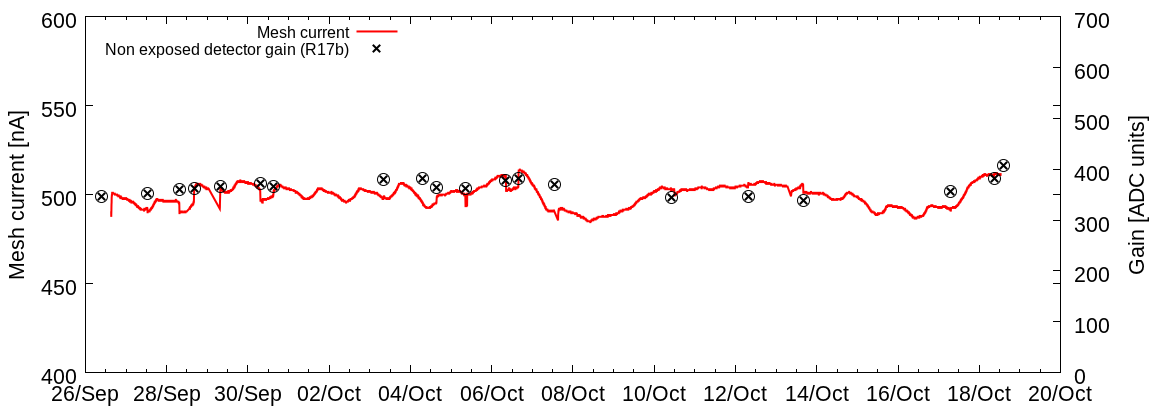}
\caption{Mesh current evolution (red curve) for a period of 21 days and an integrated charge of 918 mC. The gain control measurements with the non-exposed detector are also plotted (black circles).}
\label{xray_current}
\end{center}
\end{figure}

Measurements on the two detectors have been performed at different positions to study the relative gain before and after exposure. We used a dedicated 9-holes mask (see figure~\ref{xray_setup}) which covers the full active area of the detector in one of its axis.


\vspace{0.5cm}

\begin{figure}[htbp]
\begin{center}
\includegraphics[height=4.5cm]{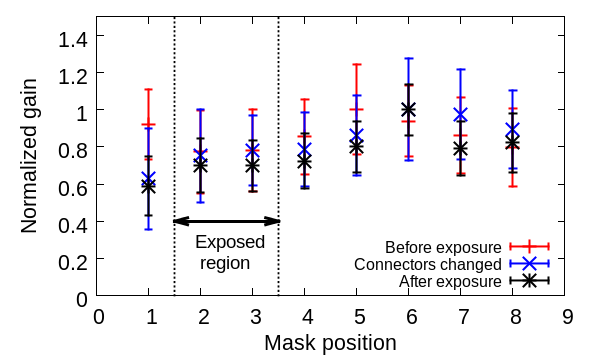}
\includegraphics[height=4.5cm]{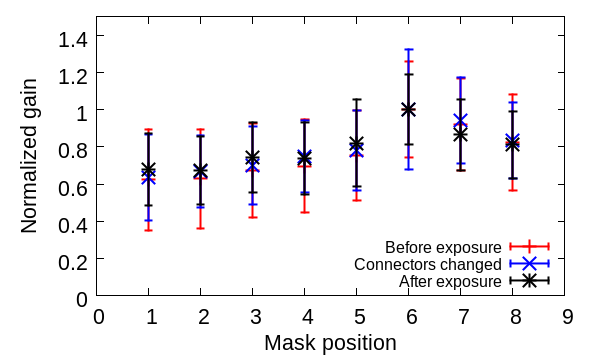}
\caption{Gain measurements as a function of position before, during, and after the irradiation period. The exposed detector R17a (left) and non-exposed detector R17b (right). The exposed region is indicated for the irradiated detector.}
\label{gain_profile}
\end{center}
\end{figure}

These measurements took place before the ageing period (26-Sep-2011), when the grounding connectors were removed (8-Oct-2011) and after the exposure (19-Oct-2011). The relative gain at each position for these three set of measurements is plotted in figure~\ref{gain_profile} and shows that the gain profile in both detectors is compatible with previous measurements. Moreover, the exposed detector region does not show a significant difference in relative gain compared to the other non-exposed regions. The X-rays irradiation had no effect on the detector response.

\subsection{Neutron exposure}

\begin{figure}[htbp]
\begin{center}
\includegraphics{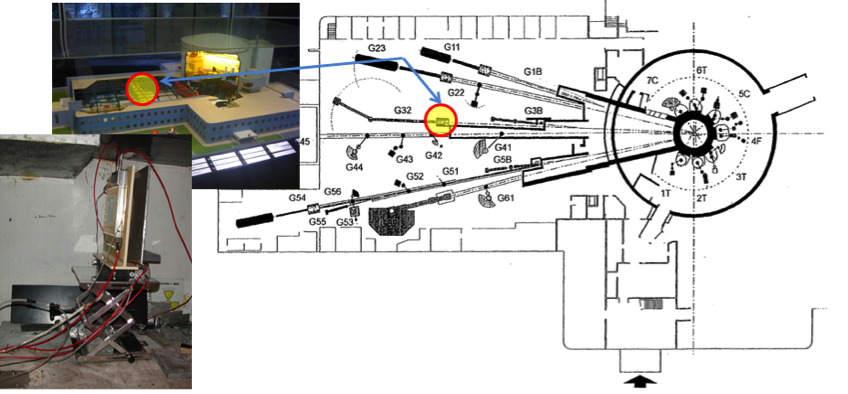}
\caption{A schematic top-view of the Orph\'ee reactor where the various cold neutron lines can be observed. The detector was installed at the cavity G32 circled in the schematic and the 3D-model by a circle. A picture of the detector as it was installed inside the cavity is also shown.}
\label{orphee}
\end{center}
\end{figure}

Neutron irradiation took place at the Orph\'ee reactor in CEA Saclay. The reactor, which operates only for research purposes, it is connected to different lines which guide different fluxes of cold neutrons produced in the reactor (see figure~\ref{orphee}). The line where the detector was installed provides a neutron beam of about 8$\times$10$^8$\,cm$^{-2}$s$^{-1}$, with energies in the range of 5-10\,meV within an area of a few cm$^2$.

\begin{figure}[htbp]
\begin{center}
\includegraphics[width=15cm]{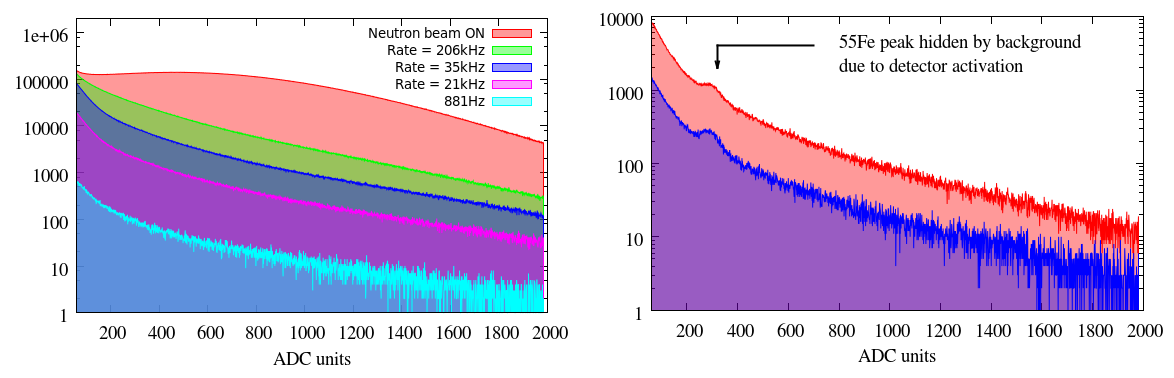}
\caption{On the left, spectra from deactivation of the detector materials after 5 minutes neutron irradiation period. On the right, $^{55}$Fe calibration over detector background due to the activation of the detector.}
\label{activation}
\end{center}
\end{figure}

The first irradiation test with neutrons lasted for a short period of 5 minutes, during which the detector materials activation reached levels that saturated the acquisition of mesh signals. The activation of the detector due to this short irradiation period was observed during several hours, the rate observed at the detector being about 1 kHz after 8 hours. The 6 keV X-rays from the $^{55}$Fe source was almost imperceptible over the detector background after this period (see figure~\ref{activation}).

\begin{figure}[htbp]
\begin{center}
\includegraphics[width=15.cm]{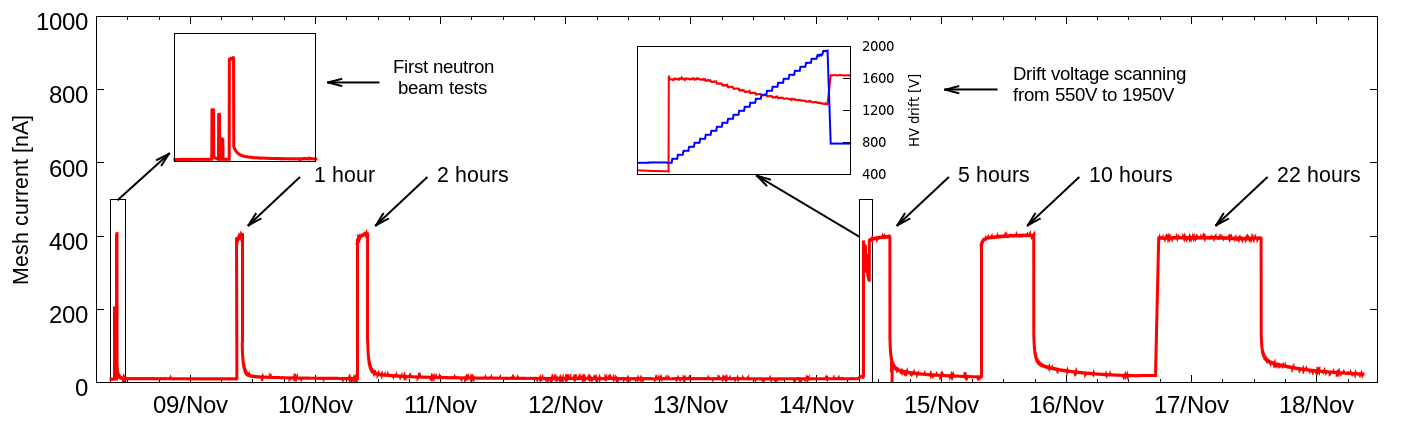}
\caption{Mesh current during the different neutron irradiation periods, during one of these periods the drift voltage was changed proving the mesh current dependency with the transparency of the detector which depends on the amplification and conversion field ratios.}
\label{neutron_current}
\end{center}
\end{figure}

The gain could not be monitored due to the activation of the detector materials during the neutron irradiation tests. Different irradiation periods were scheduled with increased time exposure (see figure~\ref{neutron_current}), The mesh current remained stable and at the same level in each of these periods.

The expected neutron flux at the CSC in ATLAS is about 3.10$^4\times$\,cm$^{-2}$s$^{-1}$. The total exposure time of the prototype R17a was more than 40 hours, accumulating a total amount of neutron flux which is equivalent to 10 years of operation of the HL-LHC with a safety factor well above~5. 

\vspace{0.1cm}

\begin{figure}[htbp]
\begin{center}
\includegraphics[height=4.5cm]{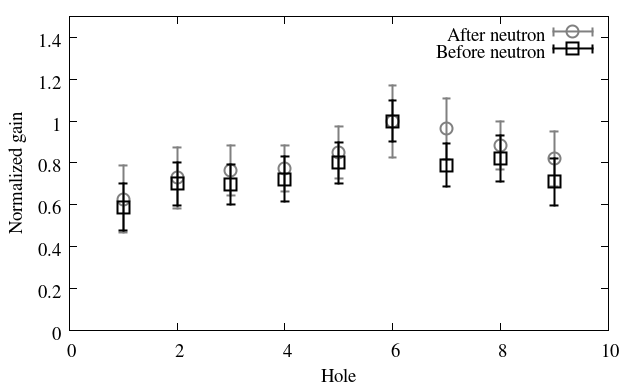}
\includegraphics[height=4.5cm]{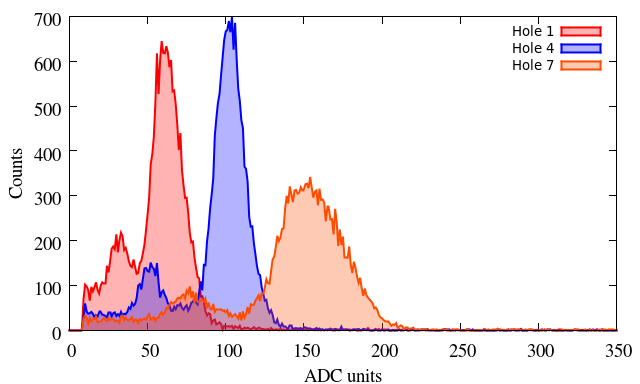}
\caption{On the left, measured gain profile before and after neutron irradiation (error bars are proportional to the energy resolution). On the right, $^{55}$Fe spectra for some of the gain points given in the left plot.}
\label{neutron_profile}
\end{center}
\end{figure}

Before and after the neutron tests the gain was monitored using the same 9-holes mask used in the tests described at the previous section. The gain profile it is compatible and the performance of the detector shows no degradation respect to the measurements before neutron irradiation (see figure~\ref{neutron_profile}).

\newpage
\subsection{Gamma exposure}

\begin{figure}[htbp]
\begin{center}
\includegraphics{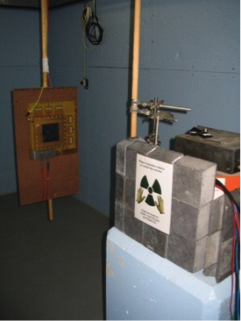}
\includegraphics{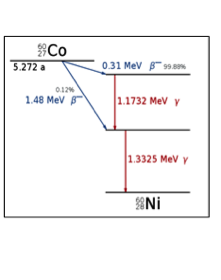}
\includegraphics{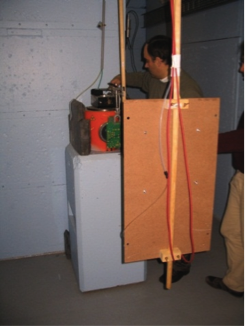}
\caption{Two pictures of the installation of the detector at the COCASE Gamma source facility, together with the gamma decay channels of a cobalt source.}
\label{cobalt}
\end{center}
\end{figure}

After proving that the detector R17a was operating properly it was installed at the COCASE gamma facility, which provides a high activity cobalt source $^{60}$Co, of about 500 mGy/h, emitting gammas at 1.17 MeV and 1.33 MeV (see figure~\ref{cobalt}). 

The highest activity for gamma at the muon spectrometer is registered in the forward CSC region~ \cite{bkg}, with a flux which is below 1.8$\times$10$^4$\,cm$^{-2}$s$^{-1}$. For 10 years of running HL-LHC, we consider a factor 5 in luminosity increase, with a security factor 3 the integrated gamma flux results to be 2.7$\times$10$^{13}$\,cm$^{-2}$.

\vspace{0.1cm}
The source was calibrated last time in summer 2005 resulting in an activity of 630 GBq~\cite{COCASE}. Considering the half-life of the $^{60}$Co to be 5.27 years, and our measurement being taken 6.5 years after we estimated the actual decay rate to be 268 GBq. The detector was placed at 50 cm from the source receiving an equivalent flux of 1.7$\times$10$^7$\,cm$^{-2}$s$^{-1}$ which should be uniformly distributed in the active area of the detector. Thus, time required to reach the expected gamma flux integrated for 10 years of HL-LHC would be of 17.9 days in COCASE.

\begin{figure}[htbp]
\begin{center}
\includegraphics[width=15cm]{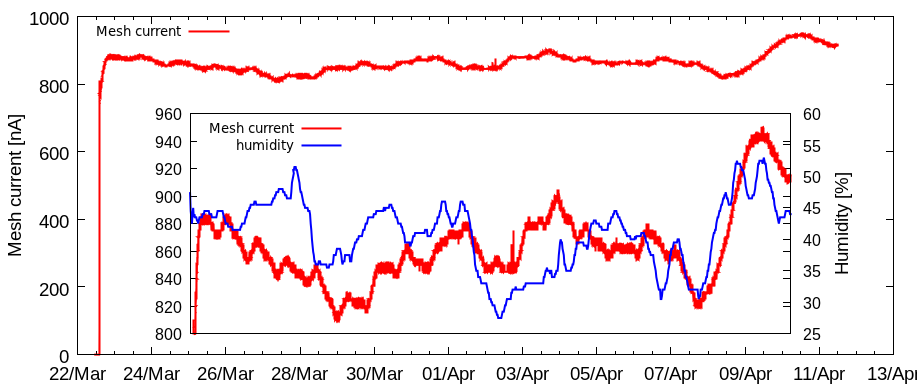}
\caption{Mesh current evolution during the gamma irradiation period, inside a zoomed plot with humidity measurements taken at COCASE facility.}
\label{gamma_current}
\end{center}
\end{figure}

The detector was irradiated between 22nd of March and 11th of April 2011, a total exposure time of 480 hours. The integrated charge during this period was 1.484 C at a mean mesh current of 858.4 nA. Figure \ref{gamma_current} shows the evolution of the mesh current which fluctuates around the mean value within a 5\%, variation which can be perfectly attributed to environmental effects (i.e. pressure and humidity variations\footnote{COCASE facility is kept at constant temperature} are also typically within 5\%).

\begin{figure}[htbp]
\begin{center}
\includegraphics[width=7.5cm]{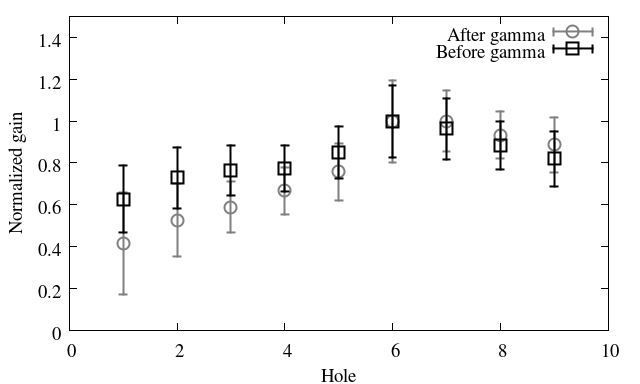}
\includegraphics[width=6.5cm]{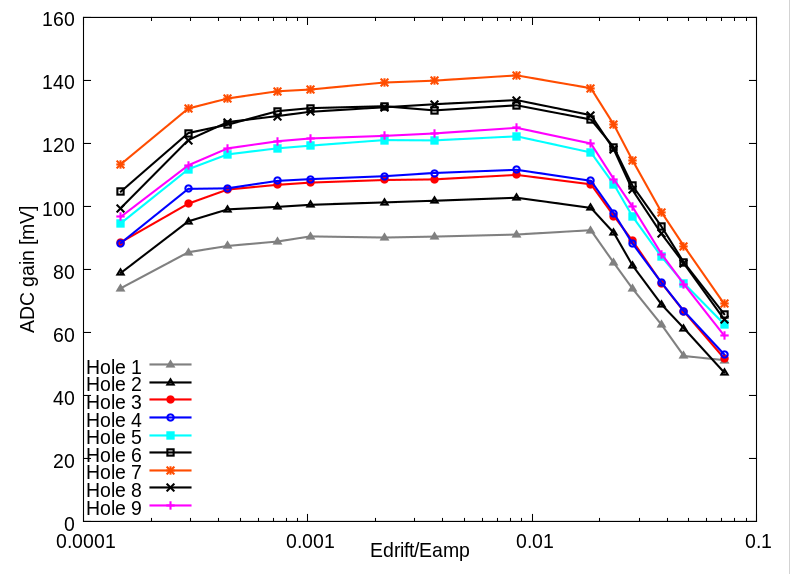}
\caption{On the left, the 9-holes mask normalized gain profile before and after gamma irradiation (error bars are proportional to the main $^{55}$Fe peak energy resolution). On the right, transparency curves at each of the mask holes.}
\label{gamma_profile}
\end{center}
\end{figure}

Gain control measurements were taken before and after exposure by using the 9-holes mask and drawing the gain profile before and after exposure, showing a reasonable reproducibility (see figure~\ref{gamma_profile}). Moreover, typical mesh transparency curves were taken at each hole showing the micromegas mesh behaved as expected.

\newpage
\subsection{Alpha exposure}

\begin{figure}[htbp]
\begin{center}
\includegraphics[width=15cm]{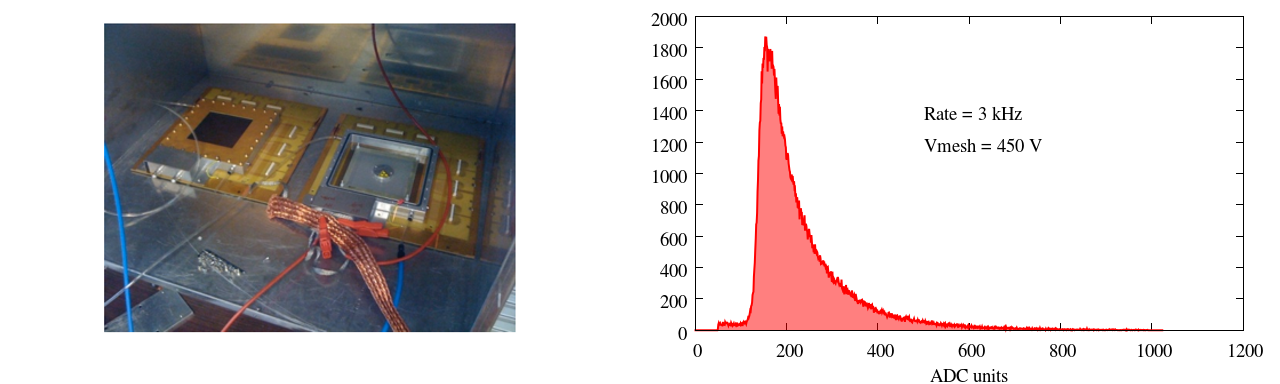}
\caption{On the left, picture of the detectors set-up. R17b on the left and R17a on the right with the $^{241}$Am source. On the right, alpha source energy deposition spectra at low gain.}
\label{alpha_setup}
\end{center}
\end{figure}

Additional tests were carried out by using a $^{241}$Am source emitting alphas at 5.64 MeV. The source was placed inside the detector chamber, just on top the metallic mesh defining the drift field. A first alpha measurement took place at low gain (around 100), allowing to determine the alpha decay rate observed by the detector (see figure~\ref{alpha_setup}). 

\begin{figure}[htbp]
\begin{center}
\includegraphics[width=15cm]{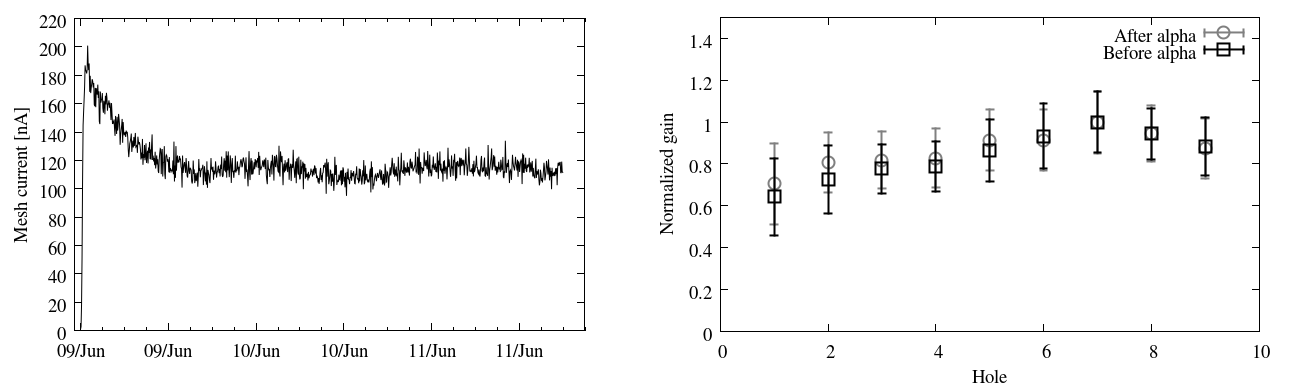}
\caption{On the left, mesh current during alpha irradiation. On the right, normalized gain profile before and after exposure.}
\label{alpha_current}
\end{center}
\end{figure}

Considering that the estimated number of primaries produced by an $^{241}$Am alpha will be between 30,000 and 60,000 for 0.5 cm conversion gap in Ar+10\%CO$_2$, the increase of gain would produce the spark conditions almost on every alpha, for gains above 5000 the critical Raether's limit \cite{raether} would be well over-passed. The gain was then increased to be about G = 7000, the alpha source stayed at this conditions for a period of 66 hours leading to a mesh current above 100 nA. After the alpha irradiation (which was localized at the center of the detector) the gain profile of the detector remained the same as it is observed by using the gain profile measurement (see figure~\ref{alpha_current}).

\newpage
\vspace{-0.4cm}
\section{Beam tests performance after irradiation}

After the completion of the above described exposures, the R17a and R17b detectors were installed in the H6 CERN-SPS beam line. The beam consists of positively charged pions at 120\,GeV/c in spills of about 10 seconds every 48\,s. The beam intensity was typically 50k pions per spill over an area of 20\,mm times 10\,mm. The micromegas detectors were recording around 800 events per spill.

\begin{figure}[htbp]
\begin{center}
\includegraphics[height=3.5cm]{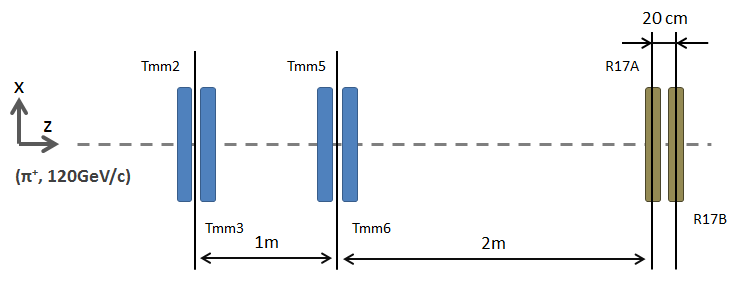}
\caption{ Test bench set-up used at the CERN pion beam facility. Beam is coming from the left and goes through reference telescope detectors Tmm$_n$. The line is shared with other detectors which were under test at the same time and or not included in these drawing for simplicity. Resistive detectors under test were placed behind the telescope. }
\label{beam_setup}
\end{center}
\end{figure}

\vspace{0.1cm}
The intention of this ultimate test was to determine any possible trace of ageing at the R17a detector by comparing the spatial resolution (SR) and efficiency obtained with respect to the R17b detector. The R17a and R17b detectors were installed in the existing micromegas telescope line, as shown in figure~\ref{beam_setup}. The micromegas telescope (Tmm2, Tmm3, Tmm5, Tmm6) is used as reference for the track reconstructions of the beam. The detectors under test were placed 2\,m downstream of Tmm6.

We estimate the intrinsic SR of the telescope detectors to be between 60-70 $\mu$m during this data taking period. Given that the detectors under test were far away from the telescope we decided to include the test detectors inside the beam track definition. In this case, we obtain an increased value for the SR of the telescope (Tmm) detectors coming from multiple scattering and extrapolation. We do not correct for this here since we are only interested in relative changes and not in the absolute SR.

\begin{figure}[htbp]
\begin{center}
\includegraphics[height=4.cm]{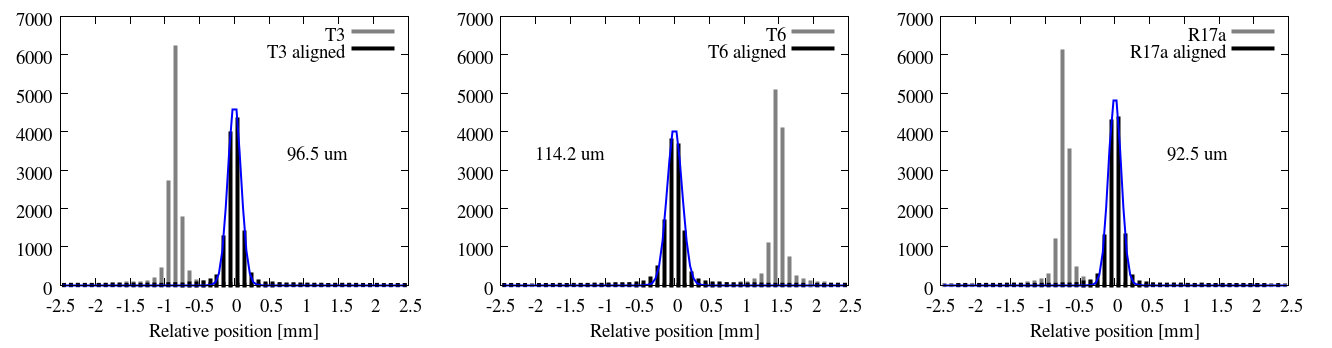}
\caption{ Relative cluster position alignment with fitted tracks. We show, before and after the alignment, the residuals of the two telescope detectors Tmm3 and Tmm5, together with the residuals of one of the resistive detectors under test, R17a.}
\label{alignement}
\end{center}
\end{figure}

\begin{figure}[htbp]
\begin{center}
\includegraphics[height=5cm]{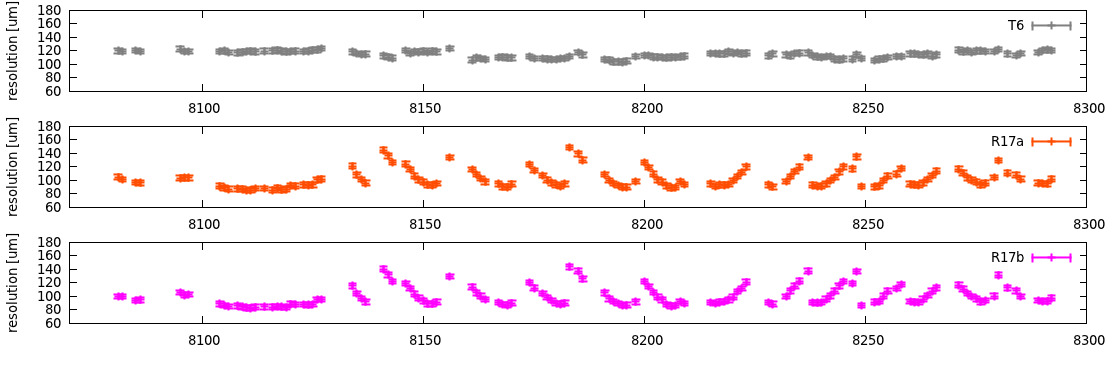}
\caption{ SR of R17a and R17b (two lower panels) and SR of one of the telescope detectors during the full beam tests period. }
\label{history}
\end{center}
\end{figure}

The data were taken within two weeks at the end of October 2012. Three different detector regions were exposed to the beam in order to compare zones which have received different types of radiation, as described in section~\ref{sc:ageing}. The regions exposed to the beam correspond to the two X-ray irradiation mask positions (left and right at about 2.5\,cm from the center) and the center of the detector. We took data at each of these regions for several values of the amplification field by varying the mesh voltage. We performed several scan repetitions at these regions by applying increasing and decreasing voltage sequences. Figure~\ref{history} shows the SR at each measurement as a function of the run number, revealing the different mesh voltage scans performed. Run numbers < 8127 correspond to the \emph{left} zone (in this region measurements were taken only at high detector gain). Run numbers between 8140 and 8208 correspond to the \emph{center} region and run numbers > 8215 correspond to the \emph{right} zone. A full scan, from low to high gains, was performed for these two last regions. The highest values for the SR observed in the plot correspond to the lowest gain values, and "vice versa".

\begin{figure}[htbp]
\begin{center}
\includegraphics[height=4.5cm]{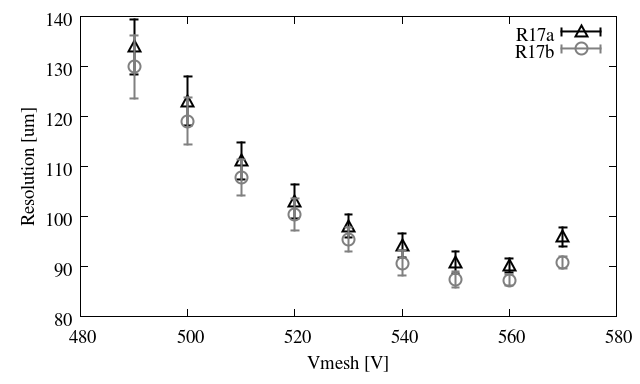}
\includegraphics[height=4.5cm]{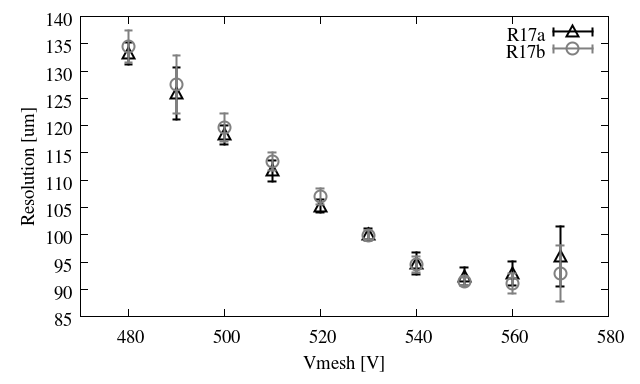}
\caption{Comparison of the SR of R17a and R17b at two regions, center and left zones respectively, as a function of the applied voltage. Error-bars represent the standard deviation of the different scans.}
\label{Resolution}
\end{center}
\end{figure}

\vspace{-0.4cm}
The improved spatial resolution for higher gains is not unexpected but emphasizes the importance of charge statistics in obtaining an accurate mean cluster charge definition at the detector. Figure~\ref{Resolution} shows the averaged SR as a function of the mesh voltage pointing to an optimum SR at about 550\,V, which corresponds to gains slightly above 10000 for the gas used at the beam tests (Ar+7\%CO$_2$).

This comparative study proofs the compatibility of the results obtained for the SR of the detectors at two different regions and between both detectors. Even if the spatial resolution obtained is not representative of the best spatial resolution that can be achieved with this type of detectors\footnote{This is due partially to the fact that the set-up was not optimized to minimize the systematics in the track determination at the resistive detectors position.}, see above, the result is good enough to describe the improvement of the spatial resolution as a function of the gain, which is validated by both detectors. It also shows that there is no difference between the SR obtained in the irradiated detector compared to the non-irradiated one.

\begin{figure}[htbp]
\begin{center}
\includegraphics[height=4.5cm]{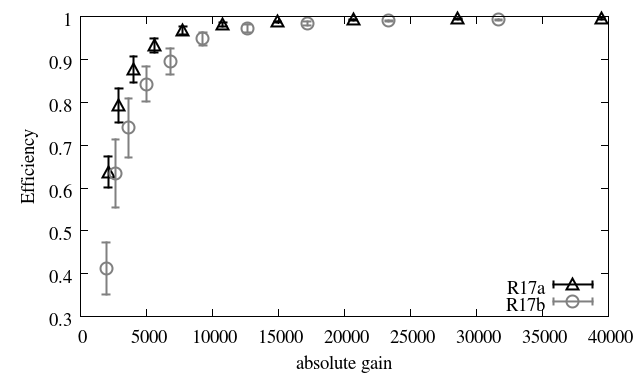}
\includegraphics[height=4.5cm]{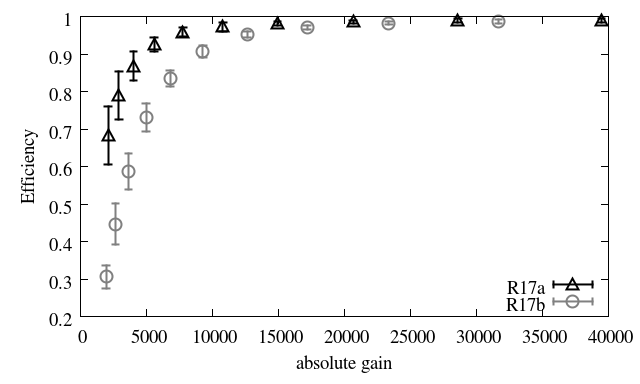}
\caption{Comparison of R17a and R17b averaged efficiencies at two different regions as a function of the absolute gain. Error-bars represent the standard deviation of the different voltage scans.}
\label{efficiency}
\end{center}
\end{figure}

\vspace{0.2cm}
Finally, the resistive detectors efficiencies of R17a and R17b are determined as the probability to observe one charge cluster when a track is observed in each of the \emph{four} telescope detectors. Figure~\ref{efficiency} shows the detection efficiency of both detectors as a function of the absolute gain. For the efficiency, both curves do not match as well as the SR. However, both detectors reach efficiencies of about 99.5\% for the highest values of the gain. It is also remarkable that the (irradiated) R17a detector reaches these values for lower gains\footnote{R17a detector showed higher gains at lower voltages along all the ageing irradiation tests.}, proving that there are no visible degradation effects in these measurements.

\section{Conclusions}

A 2D-readout resistive strip detector type was subject to different intense irradiation tests; X-ray, high energy gamma, neutrons and alphas. For each radiation type, the total charge or interactions produced at the ageing process was well above the expected levels at the HL-LHC for 10 years of operation with an additional safety factor. In addition we took these detectors to the CERN pion beam test facility. The results obtained are comparable to those obtained with a non-irradiated detector of same construction. We do not observe any degradation of the detector performance. We showed that the intense irradiation of R17a detector did not harm the performance of the detector in terms of efficiency and SR, by showing similar performances as of the R17b detector. We present it as a definitive proof that irradiations of different nature, which will be present at the HL-LHC, do not affect this new technology. The values obtained for the spatial resolution and detection efficiency are reasonably good considering that the set-up was not optimized to minimize the error on the track definition.

\vspace{0.2cm}

We have proven the robustness of the detector technology to a high level of radiation and of different nature in a relative short period of time, demanding the accumulated charge to be above the values that would be integrated in the final set-up installation in the New Small Wheel at the HL-LHC. We consider this an important step towards the consolidation of this technology for high-rate environments in long periods of time.

\vspace{0.2cm}
However, future tests should include longer period ageing at the nominal rate of the HL-LHC. In the here reported tests, giving the higher fluxes used, we cannot provide a lower limit on the mean life of these detectors by the possibility of formation of radicals which require long term polymerization process as described in~\cite{sauli}. A future test in this direction would be complementary of the results we reported.

\acknowledgments

We would like to thank to Alain Menelle for his support in the usage of the neutron beam facility; R. Chipaux and F. Daly for their advices on the installation at the gamma source in COCASE; and in general to the MAMMA collaboration and the RD51 community for their support and constructive comments for the development of this work.

\end{document}